\documentclass[11pt,titlepage,a4paper]{article}
\usepackage{bozhomac}   
\usepackage{bozhlogo}   
\usepackage{cite}   	

\title{\bfseries    \vspace*{-1.678902345in}
{\huge On the axiomatization of \\[1.22ex] ``parallel transport'' and
\\[1.22ex] its generalization
}
}

\vspace{1.7ex}

\author{
Bozhidar Z.\ Iliev
\thanks{Laboratory of Mathematical Modeling in Physics,
Institute for Nuclear Research and \mbox{Nuclear} Energy,
Bulgarian Academy of Sciences,
Boul.\ Tzarigradsko chauss\'ee~72, 1784 Sofia, Bulgaria}
\thanks{E-mail address: bozho@inrne.bas.bg}
\thanks{URL: http://theo.inrne.bas.bg/$\sim$bozho/}
}

\date{
 \vspace{2.27ex}\ShortTitle{On the axiomatization of ``parallel transport''}
								\\[0.27ex]
 \vspace{3.27ex}
\small
    \begin{tabular}{r@{$\colon\to~$}l}
 \vspace{0.09ex} Last update   & October 1, 2007    \\[0.09ex]
 \vspace{0.27ex} Produced   & \fbox{\today} \\[0.27ex]
    \end{tabular} \\[1.27ex]
\normalsize
\vspace{0.27ex}
\textsl{\bfseries
Report presented at the $8^{\text{th}}$
International Conference of Tensor Society on \\
``Differential geometry, functional and complex analysis,\\
informatics and their applications''\\
Varna, Bulgaria, 22 -- 26 August, 2005}		\\[3ex] 
    \begin{tabular}{r@{$\colon~$}l}
\normalsize\sffamily\bfseries
 \vspace{0.27ex} http://www.arXiv.org e-Print archive No. &
\normalsize\sffamily\bfseries
math-ph/0510005				\\[1.27ex]
    \end{tabular} \\[-0.27ex]
\normalsize
 \vspace{4.27ex}{\Huge \BOZHO}  \\[4.27ex]
    \begin{tabular}{r@{\hspace{0.512em}}|@{\hspace{0.512em}}l}
 \vspace{0.27ex}
&
 \vspace{0.27ex}
    \end{tabular} \\[1.27ex]
 \vspace{0.27ex}    \\[0.27ex]
}

\begin{document}        

\renewcommand{\thepage}{\roman{page}}

\renewcommand{\thefootnote}{\fnsymbol{footnote}} 
\maketitle              
\renewcommand{\thefootnote}{\arabic{footnote}}   

\tableofcontents        


\begin{abstract}

A concise discussion of the axiomatic approach to the concept of parallel
transport is presented. In particular, attention is drawn to a bijective
mapping  between the sets of connections and (axiomatically defined) parallel
transports. The transports along paths are pointed as a generalization of the
(axiomatically defined) parallel transports, which form their proper subset.
Some properties of the general transports along paths are given.

\end{abstract}

\renewcommand{\thepage}{\arabic{page}}


\section {Introduction}
\label{Introduction}

	The concept ``parallel transport'' precedes historically the one of a
``connection'' and was first clearly formulated in the
work~\cite{Levi-Civita/1917} of Levi Civita on a parallel transport of a
vector in Riemannian geometry. The connection theory was formulated
approximately during the period 1920--1949 in a series of works on
particular connections and their subsequent generalizations and has obtained
an almost complete form in 1950--1955 together with the clear formulation of
the concepts ``manifold'' and ``fibre bundle''~\cite{Lumiste-1971}. During
that period, with a few exceptional works, the `parallel transport' was
considered as a secondary concept, defined by means of the one of a
`connection'.

    The widespread approach to the concept of a ``parallel transport'' is
regarding it as a secondary one and defined on the base of the connection
theory%
~\cite{K&N-1,Sachs&Wu,Nash&Sen,Warner,Bishop&Crittenden,Yano&Kon,
Steenrod,Sulanke&Wintgen,Bruhat,Husemoller,Mishchenko,R_Hermann-1,
Greub&et_al.-1,Atiyah,Dandoloff&Zakrzewski,Tamura,Hicks,Sternberg}.
However, the opposite approach, in which the parallel transport is
axiomatically defined and from it the connection theory is constructed, is
also known%
~\cite{Lumiste-1964,Lumiste-1966,Teleman,Dombrowski,Lumiste-1971,
Mathenedia-4,Durhuus&Leinaas, Khudaverdian&Schwarz,Ostianu&et_al.,Nikolov,
Poor}
and goes back to 1949; \eg it is systematically realized in~\cite{Poor}, where
the connection theory on vector bundles is investigated.
	It seems that the earliest written accounts on this approach are
the ones due to \"U.~G.~Lumiste~\cite[sec.~2.2]{Lumiste-1964} and
C.~Teleman~\cite[chapter~IV, sec.~B.3]{Teleman} (both published
in~1964), the next essential steps being made by
P.~Dombrowski~\cite[\S~1]{Dombrowski} and W.~Poor\cite{Poor}. Besides, the
author of~\cite{Dombrowski} states that his paper is based on unpublished
lectures of prof.~Willi~Rinow~(1907--1979) in~1949; see also~\cite[p.~46]{Poor}
where the author claims that the first axiomatical definition of a parallel
transport in the tangent bundle case is given by prof.~W.~Rinow in his lectures
at the Humboldt University in~1949.

	In~\cite{bp-TP-general}, the concept of a ``parallel transport'' was
generalized to the one of a ``transport along paths.'' The relations between
both concepts ware analyzed in~\cite{bp-TP-parallelT,bp-C-TP}.

	The aim of this paper is to review some relations between the concepts
``connection on a bundle'' and ``(parallel) transport (along paths) in a
bundle''. Besides, certain elements of the theory of general transports along
paths in fibre bundles~\cite{bp-TP-general,bp-C-TP,bp-TP-morphisms} will be
presented. The set of these transports contains as a proper subset the one of
(axiomatically defined) parallel transports, which is in a bijective
correspondence with the set of connections.

	Here is the layout of the work.

	Section~\ref{Sect2} recalls the definitions of a connection on a
$C^1$ differentiable bundle and the generated by it parallel transport. The
general concept of a parallel transport on an arbitrary topological bundle is
introduced on the ground of some basic properties of these transports, .
Section~\ref{Sect3} contains elements of the general theory of transports along
paths in topological bundles; in particular, their general form is derived. In
section~\ref{Sect4} is explicitly constructed a bijection between the sets of
parallel transports and parallel transports along paths in topological bundles;
the last transports form a proper subset of the set of transports along paths.
In section~\ref{Sect5} is reestablished~\cite{bp-C-TP} the existence of a
bijection between the sets of connections and the one of parallel transports
(along paths) in $C^1$ bundles. Section~\ref{Conclusion} closes the work.
\vspace{1ex}

	The folowing notation will be used in this paper. By $(E,\pi,B)$ we
denote an arbitrary topological bundle~\cite{Husemoller,Steenrod,Sze-Tsen} with
bundle space $E$, base space $B$ and projection $\pi \colon E\to B$. The tiple
$(E,\pi,M)$ stands for a $C^1$ bundle, \ie a fibre bundle whose base and bundle
spaces are $C^1$ differentiable manifolds. The space tantent to $M$ at $x\in M$
is $T_x(M)$. An arbitrary real interval is denoted by $J$, $[a,b]$, with
$a,b\in\field[R]$ and $a\le b$, stands for a closed real interval with end
points $a$ and $b$, and $\gamma \colon J\to B$ means a path in $B$.


\section{Connections and parallel transports}
\label{Sect2}

	To begin with, we recall the most widely used definitions of a
connection and the parallel transport generated by it.

    \begin{Defn}    \label{Defn3.1}
A \emph{connection on a $C^1$ bundle} $(E,\pi,M)$ is an $n=\dim M$
dimensional distribution $\Delta^h$ on $E$ such that, for each $p\in E$
and the \emph{vertical distribution} $\Delta^v$ defined by
    \begin{equation}    \label{3.9-2}
\Delta^v\colon p \mapsto \Delta^v_p
:= T_{\imath(p)}\bigl( \pi^{-1}(\pi(p)) \bigr)
\cong T_{p}\bigl( \pi^{-1}(\pi(p)) \bigr),
    \end{equation}
with $\imath\colon\pi^{-1}(\pi(p))\to E$ being the inclusion mapping, is
fulfilled
    \begin{equation}    \label{3.9-3}
\Delta^v_p\oplus \Delta^h_p = T_p(E) ,
    \end{equation}
where
\(
\Delta^h\colon p \mapsto \Delta^h_p \subseteq T_{p}(E)
\)
and $\oplus$ is the direct sum sign. The distribution $\Delta^h$ is
called \emph{horizontal}.
    \end{Defn}

	The ($\Delta$-) horizontal lift $\bar{\gamma}$ of a path
$\gamma \colon J\to M$ is a path $\bar{\gamma} \colon J\to E$ such that
$\pi\circ\bar{\gamma}=\gamma$ and
$\dot{\bar{\gamma}}(s)\in\Delta_{\gamma(s)}^h$ for all $s\in J$, \ie the
vectors tangent to $\bar{\gamma}$ belongs to the horizontal distribution
$\Delta^h$.

    \begin{Defn}    \label{Defn3.2}
Let $\gamma\colon[\sigma,\tau]\to M$, with $\sigma,\tau\in\field[R]$ and
$\sigma\le \tau$, and
$\bar\gamma_p$ be the unique horizontal lift of $\gamma$ in $E$ passing
through $p\in\pi^{-1}(\gamma([\sigma,\tau]))$.
The \emph{parallel transport (translation, displacement)} generated by
(assigned to, defined by) a connection $\Delta^h$ is a mapping
$\Psf\colon\gamma\mapsto\Psf^\gamma$, assigning to the path $\gamma$ a mapping
    \begin{equation}    \label{3.9-4}
\Psf^\gamma\colon \pi^{-1}(\gamma(\sigma)) \to \pi^{-1}(\gamma(\tau))
\qquad \gamma\colon[\sigma,\tau]\to M
    \end{equation}
such that, for each $p\in\pi^{-1}(\gamma(\sigma))$,
    \begin{equation}    \label{3.9-5}
\Psf^\gamma(p) := \bar\gamma_p(\tau).
    \end{equation}
    \end{Defn}

	The problem of axiomatizing the concept of a parallel transport
(generated by a connection) consists in finding a set of mappings which is in a
bijective correspondence with the set of connections and with the one of
parallel transports assigned to connections.~%
\footnote{~%
A good analysis of this problem for vector bundles and linear connections on
them is contained in~\cite[chapter~2]{Poor}. %
}
Any mapping belonging to such a set can be called a parallel transport with
possible adjective(s) pointing to its origin or some its peculiarity; however,
the term ``parallel transport'' will be reserved in this work for a particular
kind of mappings described below via definition~\ref{Defn8.3}.

	The most direct way for describing what a ``parallel transport'' is, is
by axiomatizing some of the properties of the parallel transports generated by
connections.

    \begin{Prop}    \label{Prop8.2}
Let
    \begin{equation}    \label{8.10-1}
\Psf\colon\gamma \mapsto \Psf^\gamma
\colon\pi^{-1}(\gamma(\sigma))\to \pi^{-1}(\gamma(\tau))
\qquad \gamma\colon[\sigma,\tau]\to M
    \end{equation}
be the parallel transport generated by a connection on some $C^1$ bundle
$(E,\pi,M)$. The mapping $\Psf$ has the following properties:%
\\\indent\textbf{\textup{(i)}}
    The parallel transport $\Psf$ is invariant under orientation
preserving changes of the paths' parameters. Precisely, if
$\gamma\colon[\sigma,\tau]\to M$ and
$\chi\colon[\sigma',\tau']\to[\sigma,\tau]$ is an orientation preserving
 $C^1$ diffeomorphism, then
    \begin{equation}    \label{8.11}
\Psf^{\gamma\circ\chi} = \Psf^{\gamma} .
    \end{equation}
\indent\textbf{\textup{(ii)}}
    If $\gamma\colon[0,1]\to M$ and $\gamma_{\_}\colon[0,1]\to M$ is its
canonical inverse, $\gamma_{\_}(t)=\gamma(1-t)$ for $t\in[0,1]$, then
    \begin{equation}    \label{8.12}
\Psf^{\gamma_{\_}} = \bigl(\Psf^\gamma\bigr)^{-1}.
    \end{equation}
\indent\textbf{\textup{(iii)}}
    If $\gamma_1,\gamma_2\colon[0,1]\to M$, $\gamma_1(1)=\gamma_2(0)$, and
$\gamma_1\gamma_2\colon[0,1]\to M$ is their canonical product,~%
\footnote{~%
By definition, we have  $(\gamma_1\gamma_2) \colon [0,1]\to M$ and
 $(\gamma_1\gamma_2)(t)=\gamma_1(2t)$ for $t\in[0,1/2]$ and
 $(\gamma_1\gamma_2)(t)=\gamma_2(2t-1)$ for $t\in[1/2,1]$.%
}
then
    \begin{equation}    \label{8.13}
\Psf^{\gamma_1\gamma_2} = \Psf^{\gamma_2} \circ \Psf^{\gamma_1} ,
    \end{equation}
where  $\circ$ denotes composition of mappings.\\
\indent\textbf{\textup{(iv)}}
    If $\gamma_{r,x}\colon\{r\}=[r,r]\to\{x\}$ for some given $r\in\field[R]$
and $x\in M$, then
    \begin{equation}    \label{8.14}
\Psf^{\gamma_{r,x}} = \id_{\pi^{-1}(x)}
    \end{equation}
where  $\id_X$ is the identiy mapping of a set $X$.
    \end{Prop}

    \begin{Proof}
The proofs of~\eref{8.11}--\eref{8.14} can be found in a number of works, for
example
in~\cite{K&N-1,Nash&Sen,Nikolov,Lumiste-1971,Durhuus&Leinaas,
Khudaverdian&Schwarz,Lumiste-1964}
    \end{Proof}

    \begin{Defn}    \label{Defn8.3}
Let $(E,\pi,B)$ be a topological bundle. A mapping
    \begin{equation}    \label{8.11-0}
\Psf\colon\gamma \mapsto \Psf^\gamma
\colon\pi^{-1}(\gamma(\sigma))\to \pi^{-1}(\gamma(\tau))
\qquad \gamma\colon[\sigma,\tau]\to B
    \end{equation}
satisfying~\eref{8.11}--\eref{8.14} (with $B$ for $M$) will be called
\emph{(axiomatically defined) (topological) parallel transport}.
    \end{Defn}

	To justify this definition, one should prove the existence on $C^1$
bundles of bijections between the sets of parallel transports and the ones of
connections and of parallel transports generated by connection. A peculiarity
here is that, in a case of differentiable bundles, one should add some
conditions concerning the differeintiability of the parallel transports
which will be explained in section~\ref{Sect5} below. As a result of
definition~\ref{Defn3.2} and proposition~\ref{Prop8.2}, it is sufficient to be
constructed a bijection between the sets of parallel transports and
connections, which is already done in~\cite[electronic version~2 or
later]{bp-C-TP}.


\section{Transports along paths in topological fibre bundles}
\label{Sect3}

	In this section we shall describe briefly one generalization of the
parallel transport (generated by connections) which realizes one possible
generalization of the axiomatization of this concept.

    \begin{Defn}    \label{Defn8.1}
    A \emph{transport along paths} in a topological  bundle $(E,\pi,B)$ is a
mapping $I$ assigning to every path $\gamma\colon J\to B$ a mapping
$I^\gamma$, termed \emph{transport along} $\gamma$, such that
$I^\gamma\colon (s,t)\mapsto I^\gamma_{s\to t}$ where the mapping
    \begin{equation}    \label{8.1}
I^\gamma_{s\to t} \colon  \pi^{-1}(\gamma(s)) \to
\pi^{-1}(\gamma(t))
    \qquad s,t\in J,
    \end{equation}
called \emph{transport along $\gamma$ from $s$ to} $t$, has the properties:
    \begin{alignat}{2}  \label{8.2}
I^\gamma_{s\to t}\circ I^\gamma_{r\to s} &=
            I^\gamma_{r\to t} &\qquad  r,s,t&\in J
\\          \label{8.3}
I^\gamma_{s\to s} &= \id_{\pi^{-1}(\gamma(s))} & s&\in J .
    \end{alignat}
    \end{Defn}

    An analysis and various comments on this definition can be found
in~\cite{bp-TP-general,bp-TP-parallelT,bp-NF-LTP,bp-NF-D+EP}.

    As we shall see below, an important special class of transports along
paths is selected by the conditions
    \begin{alignat}{2}   \label{8.5}
I_{s\to t}^{\gamma|J'} & = I_{s\to t}^{\gamma} &\qquad &s,t\in J'
\\          \label{8.6}
I_{s\to t}^{\gamma\circ \chi} & = I_{\chi(s)\to \chi(t)}^{\gamma}
&& s,t\in J^{\prime\prime},
    \end{alignat}
where $J'\subseteq J$ is a subinterval, $\gamma|J'$ is the restriction of
$\gamma$ to $J'$, and $\chi\colon J^{\prime\prime}\to J$ is a bijection of a
real interval $J^{\prime\prime}$ onto $J$.

    Putting $r=t$ in~\eref{8.2} and using~\eref{8.3}, we see that the
mappings~\eref{8.1}  are invertible and
    \begin{equation}    \label{8.4}
(I_{s\to t}^{\gamma})^{-1} = I_{t\to s}^{\gamma} .
    \end{equation}

	For discussion and more details on the transports along paths, the
reader is referred to~\cite{bp-TP-general,bp-TP-parallelT}; in particular,
these references contain possible restrictions on them and their relations with
other differential\ndash geometric structures.

    \begin{Exmp}    \label{Exmp3.1}
Suppose the bundle $(E,\pi ,B)$ has a foliation structure~\cite{Tamura}, i.e.
on the total bundle space $E$, which now is supposed to be a manifold, to be
given a foliation $\{K_{\alpha }: K_{\alpha }\subset E$, $\alpha \in A\}$, with
$A$ being a set of indexes, which, in particular, means that~\cite{Tamura}
$K_{\alpha }\cap K_{\beta }=\varnothing$, $\alpha,\beta\in A$,
$\alpha\neq\beta$ and $\cup_{\alpha\in A} K_{\alpha }=E$. Let the foliation
$\{K_{\alpha }\}$ be such that $\pi (K_{\alpha })=B$ for all $\alpha \in A$.

	For any path $\gamma \colon J\to B$ and $u\in E$ such that
$\pi(u)\in\gamma(J)$, there is a unique lifting $\bar{\gamma}_u$ of $\gamma$
in  $E$ passing through $u$ and lying entirely in $K_{\alpha(u)}$, with
$\alpha(u)$ being the unique $\alpha(u)\in A$ such that $u\in K_{\alpha(u)}$.
It is given by
    \begin{equation}    \label{3.1}
\bar{\gamma }_{u}(s)
:= \pi ^{-1}(\gamma (s)) \cap K_{\alpha (u)},\quad s\in J.
    \end{equation}
Then one can verify that the mapping
 $K \colon \gamma\mapsto K^\gamma \colon (s,t)\mapsto K^\gamma_{s\to t}$ with
    \begin{equation}    \label{3.2}
K^{\gamma }_{s\to t}(u):=\bar{\gamma }_{u}(t) \qquad u\in \pi ^{-1}(\gamma (s)),
\quad s\in J
    \end{equation}
is a transport along paths in $(E,\pi,M)$.
    \end{Exmp}

    \begin{Exmp}    \label{Exmp3.2}
Consider the trivial bundle $(B\times G,\pr_1,B)$ where $B$ is a topological
space, $G$ is a group with multiplication
$G\times  G\ni(a,b)\mapsto a\cdot b\in G$ and $\pr_1 \colon B\times G\to B$ is
the projection on $B$. An element $u\in B\times G$ is of the form $u=(x,g)$ for
some $x\in B$ and $g\in G$ and the fibre over $x\in B$ is
$\pr_1^{-1}(x)=\{x\}\times G=\{(x,a):a\in G\}$. For $\gamma \colon J\to B$ and
$s,t\in J$, define
\(
I \colon \gamma\mapsto I^\gamma
  \colon (s,t)\mapsto I^\gamma_{s\to t}
  \colon \pr_1^{-1}(\gamma(s))\to \pr_1^{-1}(\gamma(t))
\)
by
    \begin{align}    \label{3.3}
I^\gamma_{s\to t}(u)
& = \bigl(  \gamma(t), (f(\gamma,t))^{-1} \cdot f(\gamma,s) \cdot g \bigr)
\quad \text{for } u=(\gamma(s),g)\in\pr_1^{-1}(\gamma(s))
\intertext{or by}    \label{3.3'} \tag{\ref{3.3}$^\prime$}
I^\gamma_{s\to t}(u)
& = \bigl(  \gamma(t), g\cdot f(\gamma,s)\cdot (f(\gamma,t))^{-1}  \bigr)
\quad \text{for } u=(\gamma(s),g)\in\pr_1^{-1}(\gamma(s))
    \end{align}
for some mapping $f \colon (\gamma,s)\mapsto f(\gamma,s)\in G$ for every path
$\gamma$ and $s$ in its domain. The verification  of~\eref{8.1}--\eref{8.3} is
trivial and hence $I$ is a transport along paths. In general, this transport is
not a parallel transport (see section~\ref{Sect4} below);
if $(B\times G,\pr_1,B)$ is a differentiable bundle, there does not exist a
connection for which $I_{s\to t}^{\gamma|[s,t]}$, with $s\leq t$, is the
parallel transport along the restricted path $\gamma|[s,t]$, the cause for which
is that equations like~\eref{8.11} and~\eref{8.13} do not hold for these
mappings unless some additional conditions are satisfied.
    \end{Exmp}

	The general form of the transports along paths is described by the
following
theorem.

    \begin{Thm}    \label{Thm3.1}
Let in the base $B$ of a topological fibre bundle $(E,\pi,B)$ be
given a path $\gamma : J\to B$ and, for arbitrary $s,t\in J$, be given a
mapping~\eref{8.1}. The mappings $I^{\gamma }_{s\to t}$, $s,t\in J$, define a
transport along $\gamma $ from $s$ to $t$, i.e.~\eref{8.2} and~\eref{8.3} are
satisfied, iff there exist a set $Q$ and a family of bijections
$\{F^{\gamma }_{s}:\pi ^{-1}(\gamma (s))\to Q, s\in J\}$ such that
    \begin{equation}    \label{3.6}
I^{\gamma }_{s\to t}={\bigl(}F^{\gamma }_{t}{\bigr)}^{-1}\circ
\bigl( F^{\gamma }_{s}{\bigr)},\quad s,t\in J.
    \end{equation}
    \end{Thm}

    \begin{Proof}
The theorem is a corollary of the following lemma in which one
has to put $N=J$, $Q_{s}=\pi ^{-1}(\gamma (s))$ and
$R_{s  \to t}=I^{\gamma}_{s\to t}$, $s,t\in J$.
    \end{Proof}

    \begin{Lem}    \label{Lem3.1}
Let there be given a  non-empty set $N$ and families
of equipollent sets $\{Q_{s}: s\in N\}$ and of mappings
$\{R_{s\to t}: R_{s\to t}:Q_{s}\to Q_{t},\ s,t\in N\}$. Then, the mappings
$R_{s\to t}$ satisfy the equalities
    \begin{align}    \label{3.7}
& R_{s\to t}\circ R_{r\to s}=R_{r\to t},\quad  r,s,t\in N
\\    		     \label{3.8}
& R_{s\to s}={\it id}_{Q_{s}},\quad  s\in N
    \end{align}
iff there exists an equipollent with $Q_{s}$, for some $s\in N$, set $Q$ and a
family of bijections $\{F_{s}: F_{s}:Q_{s}\to Q, s\in N\}$, such that
    \begin{equation}    \label{3.9}
  R_{s\to t}=(F_{t})^{-1}\circ (F_{s}),\quad  s,t\in N.
    \end{equation}
    \end{Lem}

    \begin{Proof}
The sufficiency is almost evident: the substitution of~\eref{3.9}
into~\eref{3.7} and~\eref{3.8} converts them into identities. Conversely,
putting $r=t$ in~\eref{3.7} and using~\eref{3.8}, we see that $R_{s\to t}$ has
an inverse mapping and
    \begin{equation}    \label{3.10}
 (R_{s\to t})^{-1}=R_{t\to s},\quad  s,t\in N
    \end{equation}
due to which, for any fixed $s_{0}\in N$, we have (see also~\eref{3.7})
\(
R_{s\to t}=R_{s_{0}\to t}\circ R_{s\to s_{0}}
={\bigl(}R_{t\to s_{0}}{\bigr)}^{-1}\circ {\bigl(}R_{s\to s_{0}}{\bigr)},
\)
i.e.~\eref{3.9} is fulfilled for $Q=Q_{s_{0}}$ and $F_{s}=R_{s\to s_{0}}$.
    \end{Proof}

	The arbitrariness in the choice of the set $Q$ and the family
$\{F_{s}\}$ in theorem~\ref{Thm3.1} is described by

    \begin{Prop}    \label{Prop3.5}
Let in a topological bundle $(E,\pi ,B)$ be given a transport $I$ along paths
with a representation~\eref{3.6} for some set $Q$ and family of bijections
$\{F^{\gamma }_{s}:\pi ^{-1}(\gamma (s))\to Q,\  s\in J\}$. Then, there exist a
set $\lindex[\mspace{-1.5mu}Q]{}{\circ}$ and family of bijections
\(
\{{\lindex[\mspace{-1mu}F]{}{\circ}}^{\gamma }_{s} :
\pi ^{-1}(\gamma (s))\to {\lindex[\mspace{-1.5mu}Q]{}{\circ}},\  s\in J\}
\) such that
    \begin{equation}    \label{3.6'}
			\tag{\ref{3.6}$^\prime$}
I^{\gamma }_{s\to t}
={\bigl(}{\lindex[\mspace{-2.2mu}F]{}{\circ}}^{\gamma }_{t}{\bigr)}^{-1}
	\circ {\bigl(}{\lindex[\mspace{-2.2mu}F]{}{\circ}}^{\gamma}_{s}{\bigr)}
	 \quad s,t\in J
    \end{equation}
iff there exists a bijection
$D^{\gamma}:{\lindex[\mspace{-1.5mu}Q]{}{\circ}}\to Q$ for which
    \begin{equation}    \label{3.11}
 F^{\gamma }_{s}=D^{\gamma }\circ {\bigl(}{\lindex[\mspace{-2.2mu}F]{}{\circ}}^{\gamma }_{s}{\bigr)}
\quad s\in J .
    \end{equation}
    \end{Prop}

    \begin{Proof}
The proposition is a consequence from the following lemma for
$N=J$, $Q_{s}=\pi^{-1}(\gamma (s))$ and $R_{s  \to t}=I^{\gamma }_{s\to t}$,
$s,t\in J$.
    \end{Proof}

    \begin{Lem}    \label{Lem3.2}
Let there be given a non-empty set $N$, a family of equipollent sets
$\{Q_{s}: s\in N\}$, which are equipollent to a set
${\lindex[\mspace{-1.5mu}Q]{}{\circ}}$, and a family of mappings
\(
\{{\lindex[\mspace{-2.2mu}F]{}{\circ}}_{s} : {
}{\lindex[\mspace{-2.2mu}F]{}{\circ}}_{s}:Q_{s}\to
{\lindex[\mspace{-1.5mu}Q]{}{\circ}}, s\in N\}.
\)
If
    \begin{equation}    \label{3.9'}
			\tag{\ref{3.9}$^\prime$}
  R_{s\to t}={\bigl(}{\lindex[\mspace{-2.2mu}F]{}{\circ}}_{t}{\bigr)}^{-1}\circ
\bigl({\lindex[\mspace{-2.2mu}F]{}{\circ}}_{s}{\bigr)} ,
    \end{equation}
then~\eref{3.9} is valid for some family of mappings
$\{F_{s}: F_{s}:Q_{s}\to Q, s\in N\}$, $Q$ being a set equipollent with
${\lindex[\mspace{-1.5mu}Q]{}{\circ}}$, if and only if there exists a
bijection $D:{\lindex[\mspace{-1.5mu}Q]{}{\circ}}\to Q$, such that
    \begin{equation}    \label{3.12}
F_{s}=D\circ ({\lindex[\mspace{-2.2mu}F]{}{\circ}}_{s}) .
    \end{equation}
    \end{Lem}

    \begin{Proof}
The sufficiency is almost evident: if~\eref{3.12} is true for some $Q$ and
$\{F_{s}\}$, then from it we find
${\lindex[\mspace{-2.2mu}F]{}{\circ}}_{s}=D^{-1}\circ F_{s}$, $s\in N$, and
substituting this result into~\eref{3.9'}, we get~\eref{3.9}. Conversely,
if~\eref{3.9} is true, then we get
\(
\bigl( F_{t}{\bigr)}^{-1}\circ \circ {\bigl(}F_{s}{\bigr)}
=\bigl(^{o} F_{t}{\bigr)}^{-1}\circ {\bigl(}
{\lindex[\mspace{-2.2mu}F]{}{\circ}}_{s}{\bigr)},
\)
$s,t\in N$,
due to~\eref{3.9'}. Hereof we see that
\(
F_{s}\circ {\bigl(}{\lindex[\mspace{-2.2mu}F]{}{\circ}}_{s}{\bigr)}^{-1}
=F_{t}\circ {\bigl(}{\lindex[\mspace{-2.2mu}F]{}{\circ}}_{t}{\bigr)}^{-1}
\)
for any $s,t\in N$, but this means that the left and right hand sides of the
last equality do not depend either on $s$ or on $t$. Hence, fixing arbitrarily
some $s_{0}\in N$ and putting
\(
D=F_{s_{0}}\circ  \circ {\bigl(}
{\lindex[\mspace{-2.2mu}F]{}{\circ}}_{s_{0}} \bigr)^{-1} :
{\lindex[\mspace{-1.5mu}Q]{}{\circ}}\to Q ,
\)
we get~\eref{3.12} from the last equality for $t=s_{0}$.
    \end{Proof}


\section
[The parallel transports as special cases of the transports along paths]
{The parallel transports as special cases of the\\ transports along paths}
\label{Sect4}

	The following theorem, whose prototype
is~\cite[theorem~3.1]{bp-TP-parallelT}, describes completely the relations
between parallel transports and transports along paths in general topological
bundles.

    \begin{Thm} \label{Thm4.1}
Let $I$ be a transport along paths in an arbitrary topological bundle
$(E,\pi,B)$ and $\gamma\colon[\sigma,\tau]\to B$. If $I$ satisfies the
conditions~\eref{8.5} and~\eref{8.6}, then the mapping
    \begin{equation}    \label{8.16}
\Isf\colon\gamma\mapsto \Isf^\gamma
:= I_{\sigma\to\tau}^{\gamma}
\colon \pi^{-1}(\gamma(\sigma)) \to \pi^{-1}(\gamma(\tau))
\qquad \gamma\colon[\sigma,\tau]\to B
    \end{equation}
is a parallel transport, \ie it possess the
properties~\eref{8.11}--\eref{8.14}, with $\Isf$ for $\Psf$.

    Conversely, suppose the mapping
    \begin{equation}    \label{8.18}
\Psf\colon\gamma\mapsto \Psf^\gamma
\colon \pi^{-1}(\gamma(\sigma)) \to \pi^{-1}(\gamma(\tau))
\qquad \gamma\colon[\sigma,\tau]\to M
    \end{equation}
is a parallel transport, \ie satisfies~\eref{8.11}--\eref{8.14}, and define
the mapping
    \begin{equation}    \label{8.19-2}
P\colon\beta\mapsto P^\beta \colon (s,t)
\mapsto P_{s\to t}^{\beta}
=
	\begin{cases}
\Psf^{\beta|[s,t]}	&\text{for } s\le t \\
\bigl( \Psf^{\beta|[t,s]} \bigr)^{-1}	&\text{for } s\ge t
	\end{cases}
\qquad \beta\colon J\to B   \quad s,t\in J .
    \end{equation}
	Then the mapping~\eref{8.19-2} is a transport along paths in
$(E,\pi,B)$, which transport satisfies the conditions~\eref{8.5}
and~\eref{8.6}, with $P$ for $I$.
    \end{Thm}

    \begin{Proof}
To prove the first part of the theorem, we define
$\tau_- \colon t\mapsto 1-t$, $\tau_1 \colon t\mapsto 2t$ and
 $\tau_2 \colon t\mapsto 2t-1$ for $t\in[0,1]$.
Applying~\eref{8.2}--\eref{8.6} and using the notation of
proposition~\ref{Prop8.2}, we get:
    \begin{align*}
\Isf^{\gamma\circ\chi}
&= I^{\gamma\circ\chi}_{\sigma\to\tau}
 = I^{\gamma}_{\chi(\sigma)\to\chi(\tau)}
 = \Isf^{\gamma}
\\
\Isf^{\gamma_-}
&= I^{\gamma\circ\tau_-}_{0\to1}
 = I^{\gamma}_{\tau_-(0)\to\tau_-(1)}
 = I^{\gamma}_{1\to0}
 = (I^{\gamma}_{0\to1})^{-1}
 = (\Isf^{\gamma})^{-1}
\\
\Isf^{\gamma_1\gamma_2}
&= I^{\gamma_1\gamma_2}_{0\to1}
 = I^{(\gamma_1\gamma_2)|[1/2,1]}_{1/2\to1} \circ
   I^{(\gamma_1\gamma_2)|[0,1/2]}_{0\to1/2}
 = I^{\gamma_2\circ\tau_2}_{1/2\to1} \circ
   I^{\gamma_1\circ\tau_1}_{0\to1/2}
 = I^{\gamma_2}_{0\to1} \circ I^{\gamma_1}_{0\to1}
 = \Isf^{\gamma_2}\circ \Isf^{\gamma_1}
\\
\Isf^{\gamma_{r,x}}
&= I^{\gamma_{r,x}}_{r\to r} = \id_{\pi^{-1}(x)}.
    \end{align*}

	To prove the second part, we first note that~\eref{8.3} and~\eref{8.6}
follow from~\eref{8.14} and~\eref{8.11}, respectively, due to~\eref{8.19-2}.
As a result of~\eref{8.19-2}, we have, \eg for $s\le t$,
\[
P_{s\to t}^{\gamma|J'}
= \Psf^{(\gamma|J')|[s,t]}
= \Psf^{\gamma|[s,t]}
= P_{s\to t}^{\gamma}
\]
by virtue of $[s,t]\subseteq J'$. At last, we shall prove~\eref{8.2} for
$r\le s\le t$; the other cases can be proved similarly. Let us fix some
orientation-preserving bijections
 $\tau' \colon [0,1]\to [r,s]$ and
 $\tau^{\prime\prime} \colon [0,1]\to [s,t]$. Then:
    \begin{multline*}
P_{s\to t}^{\beta} \circ P_{r\to s}^{\beta}
= \Psf^{\beta|[s,t]}\circ \Psf^{\beta|[r,s]}
= \Psf^{\beta\circ\tau^{\prime\prime}}\circ \Psf^{\beta\circ\tau^{\prime}}
= \Psf^{(\beta\circ\tau^{\prime})(\beta\circ\tau^{\prime\prime})}
= \Psf^{\beta\circ\tau}
= \Psf^{\beta|[r,t]}
=P_{r\to t}^{\beta}
    \end{multline*}
where
$\tau \colon a\mapsto \tau^{\prime}(2a)$  for $a\in[0,1/2]$ and
$\tau \colon a\mapsto \tau^{\prime\prime}(2a-1)$ for $a\in[1/2,1]$ ,
so that $\tau(0)=\tau'(0)=r$ and $\tau(1)=\tau^{\prime\prime}(1)=t$.
    \end{Proof}

    \begin{Defn}    \label{Defn8.4-0}
A transport $I$ along paths which has the properties~\eref{8.5}
and~\eref{8.6} will be called \emph{parallel} transport along paths.
If $I$ is a parallel transport along paths, then we say that the parallel
transport~\eref{8.16} is \emph{generated by} (\emph{defined by, assigned to})
$I$. Respectively, if $\Psf$ is a parallel transport, then we
say that the (parallel) transport along paths~\eref{8.19-2} is \emph{generated
by} (\emph{defined by, assigned to}) $\Psf$.
    \end{Defn}

	Theorem~\ref{Thm4.1} simply says that there is a bijective
correspondence between the set of parallel transports along paths and the one
of parallel transports.

    \begin{Prop}    \label{Prop4.1}
Suppose $(E,\pi,M)$ is a differentiable bundle of class $C^{m+1}$, for some
$m\in\field[N]\cup\{0\}$, and $\gamma \colon J\to M$ is a $C^1$ path. Then, in
the notation and hypotheses of theorem~\ref{Thm4.1}, if the parallel transport
$I$ along paths is of class $C^m$ in a sense that
    \begin{equation}    \label{8.17}
I_{s\to t}^{\gamma}\in
\Diff^m\bigl( \pi^{-1}(\gamma(s)), \pi^{-1}(\gamma(t)) \bigr)
\qquad \gamma\colon J\to M   \quad s,t\in J ,
    \end{equation}
then the corresponding parallel transport $\Isf$ (see~\eref{8.16}) is also of
class $C^m$,
    \begin{equation}    \label{8.15}
\Isf \colon \beta\mapsto \Isf^{\beta}\in
\Diff^m\bigl( \pi^{-1}(\beta(\sigma)), \pi^{-1}(\beta(\tau)) \bigr)
\qquad \beta\colon [\sigma,\tau]\to M   \quad s,t\in J .
    \end{equation}
    \end{Prop}

    \begin{Proof}
This result is a simple corollary of~\eref{8.16} and~\eref{8.19-2}.
    \end{Proof}


\section{Connections and (parallel) transports along paths}
\label{Sect5}

	We know from the previous sections that there exists a bijection
between the sets of parallel transports and parallel transports along paths and
that to any connection (on a $C^1$ bundle) there corresponds a parallel
transport, \viz the parallel transport generate by it. Now we shall close this
range of problems by a result stating that any parallel transport (along paths)
generates a connection on a $C^1$ bundle.

    \begin{Thm}    \label{Thm8.01}
Let $I$ be a transport along paths in a bundle $(E,\pi,M)$. Let
$\gamma\colon J\to M$ be a path and, for any $s_0\in J$ and
$p\in\pi^{-1}(\gamma(s_0))$,
the lift $\bar{\gamma}_{s_0,p}\colon J\to E$ of $\gamma$ be defined by
    \begin{equation}    \label{8.7}
\bar{\gamma}_{s_0,p}(t) = I_{s_0\to t}^{\gamma}(p) \qquad t\in J .
    \end{equation}
Suppose the transport $I$ is such that:\\
\indent
(a)
($C^1$ smoothness) The path
    \begin{subequations}    \label{8.70}
    \begin{equation}    \label{8.70a}
\text{$\bar{\gamma}_{s_0,p}$ \textup{is of class} $C^1$}
    \end{equation}
for every $C^1$ path $\gamma$, $s_0$ and $p$.
\\
\indent
(b) (Initial uniqueness)
If $\gamma_i \colon J_i\to M$, $i=1,2$, are two $C^1$ paths and for some
$s_i\in J_i$ is fulfilled $\gamma_1(s_1)=\gamma_2(s_2)$ and
$\dot\gamma_1(s_1)=\dot\gamma_2(s_2)$,
then the lifted paths $\bar{\gamma_i}_{;s_i,p}$, defined via~\eref{8.7}
with $p\in\pi^{-1}(\gamma_1(s_1))=\pi^{-1}(\gamma_2(s_2))$, have equal tangent
vectors at $p$,
    \begin{equation}    \label{8.70b}
\dot{\bar{\gamma}}_{1;s_1,p}(s_1)=\dot{\bar{\gamma}}_{2;s_2,p}(s_2).
    \end{equation}
%
\indent
(c) (Linearization)
If $\gamma_i \colon J_i\to M$, $i=1,2$, are two $C^1$ paths and for some
$s_i\in J_i$ is fulfilled $\gamma_1(s_1)=\gamma_2(s_2)$, then for every
$a_1,a_2\in\field$ there exists a $C^1$ path $\gamma_3 \colon J_3\to M$
(generally depending on $a_1$, $a_2$, $\gamma_1$ and $\gamma_2$) such that
$\gamma_3(s_3)=\gamma_1(s_1)$ ($=\gamma_2(s_2)$) for some $s_3\in J_3$ and the
vector tangent to the lifted path $\bar{\gamma}_{3;s_3,p}$, defined
via~\eref{8.7} with $p\in\pi^{-1}(\gamma_3(s_3))$, at $s_3$ is
    \begin{equation}    \label{8.70c}
\dot{\bar{\gamma}}_{3;s_3,p}(s_3)
=
a_1 \dot{\bar{\gamma}}_{1;s_1,p}(s_1) + a_2 \dot{\bar{\gamma}}_{2;s_2,p}(s_2).
    \end{equation}
    \end{subequations}

	Then
    \begin{multline}    \label{8.8}
\Delta^I\colon p\mapsto\Delta_p^I
:=\Bigl\{
    \frac{\od}{\od t}\Big|_{t=s_0}\bigl( \bar{\gamma}_{s_0,p}(t) \bigr)
\\
: \gamma\colon J\to M \text{\upshape\ is $C^1$ and injective},\ s_0\in J,\
    \gamma(s_0)=\pi(p)
  \Bigr\} \subseteq T_p(E) ,
    \end{multline}
with $p\in E$, is a distribution which is a connection on $(E,\pi,M)$, \ie
    \begin{equation}    \label{8.9}
\Delta_p^v\oplus\Delta_p^I = T_p(E) \qquad p\in E,
    \end{equation}
with $\Delta^v$ being the vertical distribution on $E$,
$\Delta_p^v=T_p(\pi^{-1}(\pi(p)))$.
    \end{Thm}

    \begin{Proof}
See~\cite[electronic version~2 or later, \S~4]{bp-C-TP}
    \end{Proof}

	In general, a parallel transport according to definition~\ref{Defn8.3}
does not define a connection on a differentiable bundle. The reason in terms of
transports along paths is that equations~\eref{8.5} and~\eref{8.6} do not
imply~\eref{8.7} and~\eref{8.70}; however,~\eref{8.7} and~\eref{8.70}
imply~\eref{8.5} and~\eref{8.6} and for that reason any connection generates a
parallel transport.  Evidently, in a differentiable bundle, a parallel
transport can generate a connection if it satisfies some differentiablility
conditions.

    \begin{Prop}    \label{Prop5.2}
If $\Psf$ is a parallel transport assigned to a connection in a $C^1$ bundle,
then, besides the properties described in proposition~\ref{Prop8.2}, it has
also the properties:  \\
\indent\textbf{\textup{(v)}}
If
\(
\bar\gamma_p \colon s\in[\sigma,\tau]\to \bar\gamma(s)
:=
\Psf^{\gamma|\sigma,s]} (p)
\)
is the lifting of $\gamma$ through $p\in\pi^{-1}(\gamma(\sigma))$ defined by
$\Psf$ and the $C^1$ paths $\gamma_i \colon [\sigma_i,\tau_i]\to M$ are such
that
 $\gamma_1(\sigma_1)=\gamma_2(\sigma_2)$ and
 $\dot\gamma_1(\sigma_1)=\dot\gamma_2(\sigma_2)$,
then
(a) the lifted path
    \begin{equation}    \label{8.15-0}
\bar\gamma_p  \text{\textup{ is of class $C^1$}}
    \end{equation}
for any $C^1$ path $\gamma$ and
(b) the lifted paths $\bar\gamma_{1;p}$ and $\bar\gamma_{2;p}$ have equal
tangent vectors at $p$,
    \begin{equation}    \label{8.15-1}
\dot{\bar\gamma}_{1;p}(\sigma_1) = \dot{\bar\gamma}_{2;p}(\sigma_2) .
    \end{equation}
\indent\textbf{\textup{(vi)}}
If $\gamma_i \colon [\sigma_i,\tau_i]\to M$, $i=1,2$, are two $C^1$ paths and
$\gamma_1(\sigma_1)=\gamma_2(\sigma_2)$, then for every $a_1,a_2\in\field$ there
exists a $C^1$ path $\gamma_3 \colon [\sigma_3,\tau_3]\to M$ such that
$\gamma_3(\sigma_3)=\gamma_1(\sigma_1)$ ($=\gamma_2(\sigma_2)$) and the vector
tangent to the lifted path
\(
\bar\gamma_{3;p} \colon t_3\in[\sigma_3,\tau_3] \to
\bar\gamma_{3;p}(t_3) := \Psf^{\gamma_3|[\sigma_3.t_3]} (p),
\)
with $p\in\pi^{-1}(\gamma_3(\sigma_3))$, at $\sigma_3$ is
    \begin{equation}    \label{8.15-2}
\dot{\bar\gamma}_{3;p}(\sigma_3)
=
a_1 \dot{\bar\gamma}_{1;p}(\sigma_2) +  a_2 \dot{\bar\gamma}_{2;p}(\sigma_2) ,
    \end{equation}
where
\(
\bar\gamma_{i;p} \colon t_i\in[\sigma_i,\tau_i] \to
\bar\gamma_{i;p}(t_i) := \Psf^{\gamma_i|[\sigma_i,t_i]} (p) .
\)
    \end{Prop}

    \begin{Proof}
The proofs of~\eref{8.15-0}--\eref{8.15-2} can be found in a number of works,
for example in~\cite{K&N-1,Nash&Sen,Lumiste-1971,Durhuus&Leinaas,
Khudaverdian&Schwarz,Lumiste-1964}
    \end{Proof}

	It can be proved~\cite[electronic version~3 or later, \S~4]{bp-C-TP}
that in $C^1$ bundle any topological parallel transport
satisfying~\eref{8.15-0}--\eref{8.15-2} defines a unique connection and
\emph{vice versa}. For this reason, such a transport may be called simply
``parallel transport''. A parallel transport defines a unique (parallel)
transport along paths satisfying~\eref{8.70} and \emph{vice versa}.


\section {Conclusion}
\label{Conclusion}

	As it was pointed above in this work, there are bijective mappings
between the sets of connections, parallel transports and  parallel transports
along paths on/in  $C^1$ differentiable fibre bundles. Thus, the theory of the
transports mentioned is only a new representation/face of the connection theory
and is a particular and equivalent realizations of the axiomatization of the
theory of parallel transports generated by connections.

	However, the parallel transports (see definition~\ref{Defn8.3}), as well
as the parallel transports along paths (see definition~\ref{Defn8.4-0}),  are
meaningful concepts on arbitrary topological bundles. Hence,  in the context of
the present paper and~\cite[chapter~\mbox{IV}, sect~8.3]{Teleman}, a connection
on such bundles can be defined as or identified with a parallel transport
(along paths).

	At last, it is worth mentioning that, in our scheme, the transports
along paths are the most general objects and that there exist transports along
paths which are not parallel transports (along paths).


\section*{Acknowledgments}

	This work was partially supported by the National Science Fund of
Bulgaria under Grant No.~F~1515/2005.


\addcontentsline{toc}{section}{References}
\bibliography{bozhopub,bozhoref}
\bibliographystyle{unsrt}
\addcontentsline{toc}{subsubsection}{This article ends at page}

\end{document}